\documentclass[conference]{IEEEtran}
\usepackage{amsfonts}
\IEEEoverridecommandlockouts

\ifCLASSINFOpdf
\else
\fi
\usepackage{epsfig}
\usepackage{graphicx}
\usepackage{subfigure}
\usepackage{psfig}
\usepackage{epsf}
\usepackage[cmex10]{amsmath}
\usepackage{booktabs}
\usepackage{fancyhdr}
\usepackage{bm}
\usepackage{color}
\usepackage{stfloats}
\usepackage{graphicx}
\usepackage{float}
\newtheorem{myTheo}{\textbf{Theorem}}   

\newtheorem{lemma}{\textbf{Lemma}}
\newtheorem{rem}{Remark} 
\begin{document}
\title{ Computation Efficiency Maximization in OFDMA-Based Mobile Edge Computing Networks}
\author{Yuhang Wu, Yuhao Wang, Fuhui Zhou, and Rose Qingyang Hu
\thanks{Manuscript received May 12, 2019; revised August 20 and September 12, 2019; accepted October 22, 2019. This work was supported in part by the National Natural Science Foundation of China (61661028 and 61701214), and in part by the Intel Corporation, and in part by the US National Science
Foundation (EARS1547312), in part by the Graduate Student Innovation and Entrepreneurship Project of Jiangxi Province (YC2018-S069), in part by the Excellent Youth Foundation of Jiangxi Province (2018ACB21012) and the Postdoctoral Science Foundation of Jiangxi Province (2017KY04). Corresponding authors: Fuhui Zhou; Yuhao Wang.}
\thanks{Yuhang Wu and Yuhao Wang are with the School of Information Engineering, Nanchang University, P. R. China, 330031. (email: may\_wu@email.ncu.edu.cn, wangyuhao@ncu.edu.cn).
 Fuhui Zhou and Rose Qingyang Hu are with Department of Electrical and Computer Engineering, Utah State University,
USA. F. Zhou  is aslo with the School of Information Engineering, Nanchang University, P. R. China, 330031. (email:  zhoufuhui@ncu.edu.cn, rose.hu@usu.edu).}
}

\maketitle
\begin{abstract}
Computation-efficient resource allocation strategies are of crucial importance in mobile edge computing networks. However, few works have focused on this issue. In this letter, weighted sum computation efficiency (CE) maximization problems are formulated in a mobile edge computing (MEC) network with orthogonal frequency division multiple access (OFDMA). Both partial offloading mode and binary offloading mode are considered. The closed-form expressions for the optimal subchannel and power allocation schemes are derived. In order to address the intractable non-convex weighted sum-of ratio problems, an efficiently iterative algorithm is proposed. Simulation results demonstrate that the CE achieved by our proposed resource allocation scheme is better than that obtained by the benchmark schemes.

\end{abstract}

\IEEEpeerreviewmaketitle
\section{Introduction}
\IEEEPARstart{T}{he}  mobile-edge computing (MEC) has emerged as a promising technology that can improve the computing capability of users \cite{M. Liu}.
In an MEC system, mobile devices are able to offload computation-intensive tasks to the MEC servers for computing. In contrast to cloud computing techniques, the MEC server is deployed close to the user \cite{Y. Mao1}. Thus, it can significantly reduce the energy consumption and  latency compared to those required by the conventional cloud-based computing systems \cite{Z. Ding}. In general, there are two ways to conduct task computing, namely, partial and binary offloading \cite{S. Bi}. 
Under the partial offloading mode, the users partition the computation task into two parts and one of them can be offloaded to the MEC server for computing. Meanwhile, the computation task is not able to be partitioned in the binary offloading mode and should be completely offloaded or only computed locally.%

Since resource allocation is of vital importance in MEC networks, there are many investigations that have focused on this area \cite{D. Xu}-\cite{F. Zhou3}.
The work in \cite{D. Xu} studied a multiuser MEC system and a resource allocation scheme aimed to minimize the energy consumption was proposed.
The authors in \cite{Fuhui Zhou} proposed two alternative algorithms to tackle the computation rate optimization problem in a wireless powered MEC system where unmanned aerial vehicle (UAV) was utilized as the MEC server.
An energy consumption minimization problem was formulated in \cite{F. Wang} for a time division multiple access (TDMA)-based wireless powered MEC system with latency constraint.
Based on the work in \cite{F. Wang}, a latency minimization problem was studied in TDMA-based MEC systems \cite{J. Ren}.
An energy-efficient non-orthogonal-multiple-access (NOMA)-based MEC system was designed in \cite{Y.Pan}. Resource allocation strategies were proposed to minimize the consumed energy.

Up to now, most of the existing works have been devoted to reducing the energy consumption, or increasing the number of computed bits, but ignored the tradeoff between the computation bits and the energy consumption.
In this letter, the computation efficiency (CE) is considered, which is a new metric that can consider both energy consumption and the computed bits. It aims to optimize the computation bits per Joule energy and describe the tradeoff between the energy consumption and the achievable computed bits \cite{F. Zhou2}. In \cite{F. Zhou2}, the CE maximization problem under the partial offloading mode was studied in TDMA-based MEC systems, and a computation-efficient resource allocation scheme was proposed. We have also investigated CE maximization problem under the max-min fairness criterion in a wireless powered MEC network \cite{F. Zhou3}, where the energy harvesting mode is practically non-linear. However, resource allocation strategies have not been fully studied for the computation efficiency maximization.

Motivated by the above mentioned facts, in our work, we design computation-efficient resource allocation strategies to achieve the maximum weighted sum computation efficiency in the OFDMA-based MEC system under both partial and binary offloading modes.
Although we have investigated CE maximization in MEC systems with TDMA in \cite{F. Zhou2}, the CE maximization problems and its solutions in the OFDMA-based multiple-user MEC system are very different from those in MEC systems with TDMA. The subchannel allocation in OFDMA systems results in more complicated mixed-integer programming problems. Moreover, both the partial and binary modes are studied.
To the best knowledge of the authors, there are no investigations that have devoted to maximizing the CE in the OFDMA-based MEC system under both partial and binary offloading modes.
We propose an efficiently iterative algorithm to tackle the formulated intractable non-convex fractional optimization problems.
Simulation results show that the CE achieved by the proposed resource allocation scheme outperforms that obtained by the benchmark schemes. Moreover, it is shown that a better performance can be obtained under the partial offloading mode than that achieved in the binary offloading mode in terms of CE. Furthermore, the simulation results elucidate that there is a tradeoff between the computed bits and the CE. In addition, the performance of the proposed CE maximization framework is compared with that of two single metric optimization frameworks. The results demonstrate that the proposed scheme outperforms other schemes in terms of the CE.

The rest of letter is organized as follows. In Section II the system model is presented. The CE maximization problems are formulated in Section III. The simulation results are shown in Section IV. Section V concludes the letter.
\section{System Model}
An OFDMA-based MEC network is considered which consists of one MEC server with single-antenna and $K$ a single antenna users denoted as ${\cal K}=\{1,2,\ldots, K\}$. The system bandwidth is partitioned into $N$ orthogonal subchannels. Let ${\cal N} = \left\{ {1,2, \ldots ,N} \right\}$ denote the available subchannels. Each has a bandwidth of $B$ Hz. It is assumed that the subchannels between the MEC server and users are block fading channels such that the subchannels remain constant during each block with duration $T$. Each user can occupy multiple subchannels. However, one subchannel can only be occupied by at most one user. The operation mode of the MEC network is stated as follows.
\subsection{Operation Mode}
\subsubsection{Partial Offloading Mode}
Under this mode, the users partition the computation task into two parts and one of them is computed locally while the other part is offloaded to the MEC server for computing.
During the computation offloading process, the transmit power is denoted as ${p_{k,n}}$ and the channel power gain of user $k$ on the $n$th subchannel can be denoted by ${h_{k,n}}$. ${\rho _{k,n}}\in \{0, 1\}$ is defined as the subchannel assignment indicator variable. Specifically, ${\rho _{k,n}}=1$ if the $n$th subchannel is occupied by the $k$th user and ${\rho _{k,n}}=0$ otherwise.
Thus, the number of computed bits can be expressed as $l_{k,n} = \rho _{k,n}TB{\log _2}( {1 + \frac{{{p_{k,n}}{h_{k,n}}}}{{N_0}}})$, where $N_0$ is the noise power at each subchannel. $T$ set as the OFDMA time block.
The energy consumed during this process for user $k$ is expressed as ${E^{off}_{k, n}} = {\rho _{k,n}}{\zeta}{p_{k,n}}T + {p_c}T$, where $p_c$ is the constant circuit power consumption for offloading process and it depends on the specific hardware implementation and the application. $\zeta$ is the amplifier coefficient.

During the local computing process, let ${C_k}$ denote the cycles required for processing one bit of input data at the central processing unit (CPU) of the $k$th user, which is assumed to be the same for all the users. The local computation time is the entire block $T$. The local computing frequency of the $k$th user is denoted as $f_k$. Therefore, the computed  bits of the $k$th user is $l_k^{local} = {T{f_k}}/{C_k}$. The energy consumed in local computing is modeled as $E_k^{local} = {\varepsilon _k}f_k^3T$, where $\varepsilon_k$ denotes the coefficient depend on chip architecture at the user $k$ \cite{Fuhui Zhou}. The CE is the ratio of the computation bits both in local computing and offloading process to the consumed energy during the entire block $T$ \cite{F. Zhou2}.
According to the definition of CE in \cite{F. Zhou2} and \cite{F. Zhou3}, the CE of the $k$th user is expressed as
\begin{align}\label{27}\
\eta_k=\frac{R_k}{P_k}=\frac{\sum\nolimits_{n = 1}^N {{\rho _{k,n}}B}{\log _2}\left( {1 + \frac{{p_{k,n}}{h_{k,n}}}{N_0}} \right) + \frac{{f_k}}{{C_k}}}{\sum\nolimits_{n = 1}^N {\rho _{k,n}}{\zeta}{p_{k,n}} + {\varepsilon _k}f_k^3+ {p_c}}
\end{align}

\subsubsection{Binary Offloading Mode}
In this mode, the computation task should be completely offloading to the MEC server or only computed locally. This mode is appropriate to the scenario where the computation task of users is indivisible. Let $\mu_k \in \{0,1\}, \forall k \in \cal K$ denote the operational mode selection factor, namely, $\mu_k=1$ means that user $k$ chooses to perform task offloading mode, and $\mu_k=0$ represents the local computation mode is chosen  in user $k$. Accordingly, under binary offloading mode, the CE of each user is expressed as $\eta_k={{R}_k}/{P_k}$, where ${R_k=\mu_k\sum\nolimits_{n = 1}^N {{\rho _{k,n}}}B{\log _2}( {1 + \frac{{p_{k,n}}{h_{k,n}}}{N_0}} ) + (1-\mu_k)\frac{{f_k}}{{C_k}}}$ and ${P_k=\mu_k\sum\nolimits_{n = 1}^N {\rho _{k,n}}{\zeta}{p_{k,n}} + (1-\mu_k){\varepsilon _k}f_k^3+ {p_c}}$, respectively.
\section{Problem Formulation}
\subsection{CE Maximization Under The Partial Offloading Mode}
\subsubsection{Problem Formulation}

\begin{figure*}[ht]
\begin{small}
 \begin{align}\label{27}\
{\cal L} = \sum\limits_{k = 1}^K {{\lambda _k}} \left( {{\omega _k}{\tilde R_k} - {\beta _k}{\tilde P_k}} \right)+ \sum\limits_{k = 1}^K {{\upsilon _k}} \left( {f_k^{\max } - {f_k}} \right)  + \sum\limits_{k = 1}^K {{\alpha _k}} \left( {R_k^{th} - \tilde R{_k}} \right)  + \sum\limits_{k = 1}^K {{\varsigma _k}} \left( {P_k^{th} - \tilde P{_k}} \right) + \sum\limits_{n = 1}^N {{\xi _n}\left( {1 - \sum\limits_{k = 1}^K {{\tilde\rho _{k,n}}} } \right)}\tag{6}
\end{align}
\end{small}
\hrulefill
\end{figure*}

Under this mode, the weighted sum CE maximization problem is formulated as
 \begin{subequations}
 \begin{align}\label{27}\
{{\mathbf{P}}_1}:&\mathop {\max }\limits_{{\rho _{k,n}},{p_{k,n}},{f_k}} \sum\limits_{k = 1}^K {{\omega _k}{\eta _k}} \\
{\rm{s}}{\rm{.t}}{\rm{.}}\;&C1\;:\;{R_k} \ge R_k^{th},\forall k,\\
&C2\;:\;{P_k} \le P_k^{th},\forall k,\\
&C3\;:\;0 \le {f_k} \le f_k^{\max },\forall k,\\
&C4\;:\;\sum\nolimits_{k = 1}^K {{\rho _{k,n}}}  \le 1,\forall n,\\
&C5\;:\;{\rho _{k,n}} \in \left\{ {0,1} \right\},\forall n,\forall k,
\end{align}
\end{subequations}
where $\omega _k>0$ indicates the weight of the $k$th user and it gives the relative priority of each user. The constraint $C$1 ensures that the minimum required computation bits of each user can be satisfied. The constraint $C$2 is the total power consumption constraint, and $P_k^{th}$ denotes the maximum available power. The constraint $C$3 is the CPU frequency constraint, where $f_k^{\max}$ denotes the maximum CPU frequency of the $k$th user. The constraints $C$4 and $C$5 ensures each subchannel can only exist at most one user.

$\mathbf{P}_1$ is a challenging non-convex problem since the objective function has the sum-of-ratios structure. Moreover, the existence of coupling relationship among different variables makes the problem more complicated. In order to solve $\mathbf{P}_1$, an auxiliary variable ${z_{k,n}}$ is introduced.
The subchannel assignment indicator variable is relaxed into a continuous variable, namely, ${\tilde \rho _{k,n}}\in[0, 1]$. Let ${z_{k,n}} =  {{\tilde\rho _{k,n}}} {p_{k,n}}$, by substituting ${p_{k,n}} = {z_{k,n}}/{{\tilde\rho _{k,n}}}$ and applying a parametric algorithm to transform $P_1$. By introducing an auxiliary parameter $\beta_k$, the equivalent problem is given as
\begin{subequations}
 \begin{align}\label{27}\
\mathbf{P}_{{2}}:\ \ &\mathop {\max }\limits_{{\tilde\rho _{k,n}},{z_{k,n}},{f_k},{\beta _k}} \sum\nolimits_{k = 1}^K {{\beta _k}}\\
\text{s.t.}\ &\ \beta _k \le \frac{\omega_k {{\tilde R_k}}}{{\tilde P_k}}, \forall k.\\
&\ \tilde{R}_k \ge {R_k^{th}},\forall k,\\
&\ \tilde{P}_k \le P_k^{th},\forall k,\\
&\  C3,\ \sum\nolimits_{k = 1}^K  {{ \tilde\rho _{k,n}}}  \le 1,\forall n,\\
&\ 0\le{ \tilde\rho _{k,n}}\le1,\ \forall n,\forall k,
\end{align}
\end{subequations}
where $\tilde{R}_k=\sum\nolimits_{n = 1}^N {{\tilde\rho _{k,n}}B}{\log _2}\left( {1 + {{p_{k,n}}{h_{k,n}}}/{N_0}} \right) + {{f_k}}/{{C_k}}$ and $\tilde{P}_k=\sum\nolimits_{n = 1}^N {\tilde\rho _{k,n}}{\zeta}{p_{k,n}} + {\varepsilon _k}f_k^3+ {p_c}$. Furthermore, $\textbf{P}_2$ can be transformed into an equivalent optimization problem with an objective function in subtractive form via Lemma 1.

\begin{lemma}
If ${(\bm{{\tilde\rho^* _{k,n}},{z^*_{k,n}},{f^*_k},{\tilde\eta^*_k}})}$ is the optimal solution to $\mathbf{P_2}$, then there exist ${\bm{\lambda^*}=(\lambda_1,\cdots,\lambda_k)}$ and ${\bm{\beta^*}=(\beta_1,\cdots,\beta_k)}$ such that the optimal solution to the following problem is ${(\bm{{\tilde\rho^* _{k,n}},{z^*_{k,n}},{f^*_k}})}$ for the given $\bm{(\lambda, \beta)}$, i.e. $\bm{\lambda=\lambda^*}$ and $ \bm{\beta=\beta^*}$.
\begin{subequations}
 \begin{align}\label{27}\
&\mathbf{P}_{{3}}:\mathop {\max }\limits_{{\tilde\rho _{k,n}},{z_{k,n}},{f_k}}\sum\limits_{k = 1}^K {{\lambda _k}} \left( {{\omega _k}{\tilde R_k} - {\beta _k}{\tilde P_k}} \right)\\
\text{s.t.}&\ \ (3c)-(3f).
\end{align}
\end{subequations}
Besides, ${(\bm{{\tilde\rho^* _{k,n}},{z^*_{k,n}},{f^*_k},{\tilde\eta^*_k}})}$ also satisfies the following equations for  $\bm{(\lambda, \beta)}$:
\begin{align}\label{27}\
\lambda_k=\frac{1}{\tilde{P_k}}, \forall k, \ \ {\omega_k}{\tilde R_k}-\beta_k{\tilde P_k}=0, \forall k.
\end{align}
\end{lemma}
\begin{IEEEproof}
Please refer to \cite{Y. Jong} for the proof.
\end{IEEEproof}

The convex of the $\mathbf{P}_3$ is easy to prove. Therefore, the optimal solutions can be readily obtained by using the dual method for the given $\bm{(\lambda, \beta)}$. Then an efficient approach to updating $\bm{(\lambda, \beta)}$ is developed. By rearranging terms, the Lagrangian of $\mathbf{P}_3$ is written as (6),

where $\upsilon _k, \alpha _k, \varsigma _k$ and $\xi _n$ are non-negative Lagrange multipliers for the corresponding constraints.
\subsubsection{Optimal Solutions}
Let $p^{opt}_{k,n}$ and $f^{opt}_{k}$ denote the optimal transmit power and local computation frequency of the $k$th user, respectively.
\begin{myTheo}
The optimal transmit power and local computation frequency of the $k$th user for the weighted sum CE maximization are given as
\begin{align}\label{27}\
\setcounter{equation}{6}
p^{opt}_{k,n}& = {\left[ {\frac{{\left( {{\lambda _k}{\omega _k} - {\alpha _k}} \right) B}}{{\ln 2\zeta \left( {{\lambda _k}{\beta _k} + {\varsigma _k}} \right)}} - \frac{{ N_0}}{{h_{k,n}}}} \right]^ + },\\
f_k^{opt} &= \sqrt {{{\left[ {\frac{{\frac{{\left( {{\lambda _k}{\omega _k} - {\alpha _k}} \right)}}{{{C_k}}} - {\upsilon _k}}}{{3\left( {{\lambda _k}{\beta _k} + {\varsigma _k}} \right){\varepsilon _k}}}} \right]}^ + }},
\end{align}
where $[a]^+$ represents the maximum between $a$ and 0.
\end{myTheo}
\begin{IEEEproof}
Differentiate the Lagrangian (6) with respect to $z_{k,n}$, and let it equal 0. Since  $z_{k,n}\ge 0$, the optimal $z^{opt}_{k,n}$ is obtained as ${{z^{opt}_{k,n}} = \left[\frac{{\left( {{\lambda _k}{\omega _k} - {\alpha _k}} \right){{\tilde\rho _{k,n}}} B}}{{\ln 2\zeta\left( {{\lambda _k}{\beta _k} + {\varsigma _k}} \right)}} - \frac{{\tilde\rho _{k,n}}N_0}{{h_{k,n}}}\right]^+}$. Since ${p_{k,n}} =\frac {z_{k,n}}{{\tilde\rho _{k,n}}}$, the optimal transmit power of the $k$th user can be obtained. Similarly, the optimal local computation frequency of the $k$th user can be obtained from the derivations of the Lagrange function (6) with respect to $f_{k,n}$. The proof for \textbf{Theorem 1} is completed.
\end{IEEEproof}
\begin{rem}
Note that $p^{opt}_{k,n}$ increases with the subchannel gain $h_{k,n}$. This suggests that the user with a higher subchannel gain should allocate more power to the  offloading process for the purpose of obtaining the maximum weighted sum CE. Besides, the subchannel gain needs to satisfy $h_{k,n}\geq[N_0\ln2\zeta(\lambda_k\beta_k+\varsigma _k)]/(\lambda_k\omega_k-\alpha_k)B$. Otherwise,  the optimal power $p^{opt}_{k,n}$ becomes zero, which means the user  can only perform local computation.
\end{rem}

According to (6), the optimal subchannel allocation $\rho^{opt}_{k,n}$ is described in \textbf{Theorem 2}.
\begin{myTheo}
The optimal subchannel allocation of the OFDMA-based MEC system is obtained as
\begin{small}
\begin{align}\label{27}\
i\left( k \right) = \arg \mathop {\max }\limits_{1 \le k \le K}{H_{k,n}}, \rho _{i\left( k \right),n}^{opt} = 1, \rho _{k,n}^{opt} = 0,\forall k \ne i\left( k \right).
\end{align}
\end{small}
\textbf{Theorem 2} indicates that the $n$th subchannel should be allocated to the user with the largest ${H_{k,n}}, k\in \cal{K}$. ${H_{k,n}}$ is given by (11).
\end{myTheo}
\begin{IEEEproof}
Taking partial derivative of (6) and applying the Karush-Kuhn-Tucker (KKT) conditions, one has
 \begin{align}\label{27}\
\frac{{\partial {\cal L}\left( {{\tilde\rho _{k,n}},{z_{k,n}},{f_k}}\right)}}{{\partial {\tilde\rho _{k,n}}}}\left\{ \begin{array}{l}
 < 0,\ \ \tilde\rho _{k,n}^{opt} = 0,\\
 = 0,\ \ 0 < \tilde\rho _{k,n}^{opt} < 1,\\
 > 0,\ \ \tilde\rho _{k,n}^{opt} = 1.
\end{array} \right.{\rm{     }}\forall k,n.
\end{align}
And the optimal subchannel allocation can be given as
\begin{align}\label{27}\ \notag
{H_{k,n}} = &\left( {{\lambda _k}{\omega _k} - {\alpha _k}} \right)B\left( {{\log }_2}\left( {1 + \frac{{{z_{k,n}}{h_{k,n}}}}{{N_0}}} \right) -\right.\\ &\left. \frac{{{z_{k,n}}{h_{k,n}}}}{{\ln 2\left( {N_0 + {z_{k,n}}{h_{k,n}}} \right)}} \right) ,\\
\rho _{k,n}^{opt} =& \left\{ \begin{array}{l}
0,\ {H_{k,n}} < {\xi _n}\\
1,\ {H_{k,n}} > {\xi _n}
\end{array} \right.,\forall k,n,
\end{align}
where $H_{k,n}, \forall k,n$ is the channel allocation indicator obtained by the above derivation process. It plays a vital role in finding the optimal subchannel allocation and power allocation. It reflects the tradeoff between the achievable computed bits and the energy consumption.
The proof for \textbf{Theorem 2} is completed.
\end{IEEEproof}

The subgradient method can be applied to obtain all the dual variables \cite{S. Boyd}. Finally, an efficient approach to attain the optimal CE under partial offloading mode is presented in \textbf{Algorithm 1}.
\subsubsection{Complexity Analysis}
The complexity analysis of Algorithm 1 is presented as follows. Firstly, the complexity of $\tilde\rho_{k,n},\ p_{k,n}$ and $f_k$ , $\forall k$, linearly increase with  $KN$, where $K$ and $N$ are the number of users and the number of subchannels, respectively. Secondly, since the number of dual variables is $4K+N$, the subgradient method has the complexity in $O(K^2N^2)$. Finally, the complexity of $\lambda_k$ and $\beta_k$ is independent of $K$ \cite{Y. Jong}. Thus, for the proposed algorithm, the total complexity is $O(K^3N^2)$.
\begin{table}[htbp]
\begin{center}
\begin{tabular}{lcl}
\\\toprule
$\textbf{Algorithm 1}$: Computation-efficiency resource allocation algorithm \\for OFDMA-based MEC networks\\ \midrule
\  1: Initialize the algorithm accuracy indicator $e$, $\bm\lambda$ and $\bm\beta$,\\
\ \ \ \ and set $i=0$;\\
\  2: \textbf{repeat}\\
\  3: \ \ Initialize $({\upsilon _k},{\alpha _k},{\varsigma _k},{\xi _n})$;\\
\  4: \ \ \textbf{repeat}\\
\  5: \ \ \ \ Obtain the optimal transmit power $p^{opt}_{k,n}$, local computation\\
 \ \ \ \ \ \ \ \ capacity $f^{opt}_{k,n}\ $ and the optimal subchannel allocation $\rho _{k,n}^{opt}$\\
 \ \ \ \ \ \ \ \ from (7), (8) and (9) respectively;\\
\  6: \ \ \ \ Update the Lagrange variables $({\upsilon _k},{\alpha _k},{\varsigma _k},{\xi _n})$ by \\
\ \ \ \ \ \ \ \ ${\upsilon _k}(i+1)={\upsilon _k}(i)+\Delta{\upsilon _k}(f^{max}_k -f^*_k)$,\\
 \ \ \ \ \ \ \ \ ${\alpha_k}(i+1)={\alpha _k}(i)+\Delta{\alpha _k}( R^{th}_k-\tilde R^*_k)$,\\
 \ \ \ \ \ \ \ \ ${\varsigma _k}(i+1)={\varsigma _k}(i)+\Delta{\varsigma _k}(P^{th}_k-\tilde P^*_k)$,\\
 \ \ \ \ \ \ \ \ ${\xi _n}(i+1)={\xi _n}(i)+\Delta{\xi _n}(1-\sum\limits_{k}\tilde\rho_{k,n})$;\\
\  7: \ \ \textbf{until} $({\upsilon _k},{\alpha _k},{\varsigma _k}, {\xi _n})$ converges;\\
\  8: \ \ \ Update $\bm\lambda$ and $\bm\beta$ by (5); \\
\ 9: \ \ \ $i=i+1$;\\
10: \textbf{until} $p^{opt}_{k,n}$, $f^{opt}_{k,n}$ and $\rho _{k,n}^{opt}$ converges;\\
11: Output the optimal computation efficiency.\\
\bottomrule
\end{tabular}
\end{center}
\end{table}
\subsection{CE Maximization Under The Binary Offloading Mode}
\subsubsection{Problem Formulation}
In this mode, the optimization problem is formulated as
\begin{subequations}
\begin{align}\label{27}\
\mathbf{P}_{{4}}:&\mathop {\max }\limits_{\scriptstyle{\rho _{k,n}},{p_{k,n}},
\scriptstyle{f_k},{\mu _k}\hfill} \sum\limits_{k = 1}^K\omega_k\eta_k\\
s.t.\ &R_k \ge {R_k^{th}},\forall k \in \cal K,\\
&P_k \le P_k^{th},\forall k \in \cal K,\\
&C3-C5,\\
&{\mu_k} \in \left\{ {0,1} \right\},\forall k \in \cal K,
 \end{align}
\end{subequations}
where $(13\rm{b})$ stands for the minimum computed bits requirement and $(13\rm{c})$ denotes the maximum available power for binary offloading mode, respectively. The constraint $(13\rm{e})$ indicates that each user can either only offload tasks or only perform local computing.

\subsubsection{Optimal Solutions}
It can be seen that $\mathbf{P}_4$ is similar to $\mathbf{P}_1$. Thus, $\mathbf{P}_4$ can adopt the same method for solving $\mathbf{P}_1$, for a given $\mu_k$. Moreover, $\mu_k$ is relaxed as a continuous variable, $\tilde\mu_k\in[0,1]$. The optimal solution for the operation mode selection factor can be obtained by Theorem 3.

\begin{myTheo}
The operation mode selection factor of the CE maximization problem under the binary computation offloading mode can be given as
\begin{subequations}
 \begin{align}\label{27}\
\ {\mu _k}=& \left\{ \begin{array}{l}
0,\ {F_k^1} < {F_k^2}\\
1,\ {F_k^1} \ge {F_k^2}
\end{array}, \right.\\
{F_k^1}=& \left( {{\psi_k }{\omega _k} - {\vartheta _k}} \right)\sum\limits_{n = 1}^N {{\rho _{k,n}}} B{\log _2}\left( {1 + \frac{{{x_{k,n}}{h_{k,n}}}}{{{\rho _{k,n}}N_0}}} \right)\\\notag
&- \left( {{\psi_k }{\phi _k} + {\chi _k}} \right)\left( {\sum\limits_{n = 1}^N {{x_{k,n}}}  + {p_c}} \right),\\
{F_k^2}=& \left( {{\psi_k }{\omega _k} - {\vartheta _k}} \right)\frac{{{f_k}}}{{{C_k}}} - \left( {{\psi_k }{\phi _k} + {\chi _k}} \right)\left( {{\varepsilon _k}f_k^3 + {p_c}} \right),
\end{align}
\end{subequations}
where $\vartheta _k$ and $\chi _k$ are the non-negative Lagrange factors for the constraints $(13\rm{b})$ and $(13\rm{c})$, respectively. Moreover, $\phi_k$ and $\psi_k$ are the auxiliary variable introduced according to $\textbf{Lemma 1}$.
\end{myTheo}
\begin{IEEEproof}
It is easy to prove $\textbf{Theorem 3}$ from the derivations of the Lagrangian with respect to $\mu_k$. Due to space limit, the detail is omitted .
\end{IEEEproof}

The optimal transmit power $p^{opt}_{k,n}$, local computation capacity $f^{opt}_{k,n}$ and the optimal subchannel allocation $\rho _{k,n}^{opt}$ can be similarly obtained by $\textbf{Theorem 1}$ and $\textbf{Theorem 2}$.

\section{Simulation Results}
In this section, simulation results are presented to investigate the performance of our proposed resource allocation schemes. The following parameter settings are considered. The system bandwidth is $B=2MHz$, $T$ is set as $1s$, the number of subchannel $N$ is set as $4$ and the number of users is $2$ \cite{S. Bi}. The channel fading is Rayleigh.The number of cycles for processing each bit is set to be $C_k=10^3$ cycles per bit, and the chip CE is $\varepsilon_k=1 \times 10^{-24}$. The weights for all users are set to be 1, i.e., $\omega_k=1, \forall k$. The transmit circuit power consumptions are the same for all users, $p_c=50$ mW. The amplifier coefficient $\zeta$ is set to $3$. Four baseline schemes are also evaluated for comparison which are stated as follows: 1) Offloading only; 2) Local computing only; 3) Maximizing computation bits (CB) scheme: the CB maximization; 4) Minimizing energy consumption (EC) scheme: the EC minimization.

Fig. 1 illustrates the convergence of the proposed algorithm, i.e., the CE obtained by different computing modes versus the number of iterations. It is observed that the convergence rate of the proposed algorithm is fast. It can converge to a constant with only a small number of iterations. Thus, the high efficiency of the proposed algorithm is verified. Besides, the proposed algorithm has a better performance than the other schemes in terms of CE.

From Fig. 2, it can be seen that the CE of Algorithm 1 first increases with the transmit power. Then, when the transmit power is high enough, the CE reaches the saturation point. This is because the number of computed bits increases faster than the consumed energy when a small maximum available transmit power is adopted. But when the maximum transmit power is high enough, the user chooses the optimal transmit power that maximizes the sum weighted CE instead of using the maximum transmit power. Since more available power allows users to offloading more computation data to the MEC server or processing them locally, which significantly improves the CE. 
Moreover, we can observe that the CE of our proposed schemes is larger than those of other three benchmark schemes since the proposed scheme has more flexibility in allocating resources between computation offloading or local computation. It is also shown that the CE obtained under the partial offloading scheme is superior to that achieved under the CB maximization scheme. This is due to the CB maximization scheme consumes all the available energy to obtain the maximum achievable computation bits. Since the computation bits and the consumed energy are simultaneously increased, the CE may be not increased.
\begin{figure*}[htbp]
\vspace{-2em}
\centering
\subfigure[]{
\begin{minipage}{2.25 in}
\centering
\includegraphics[width=2.3 in]{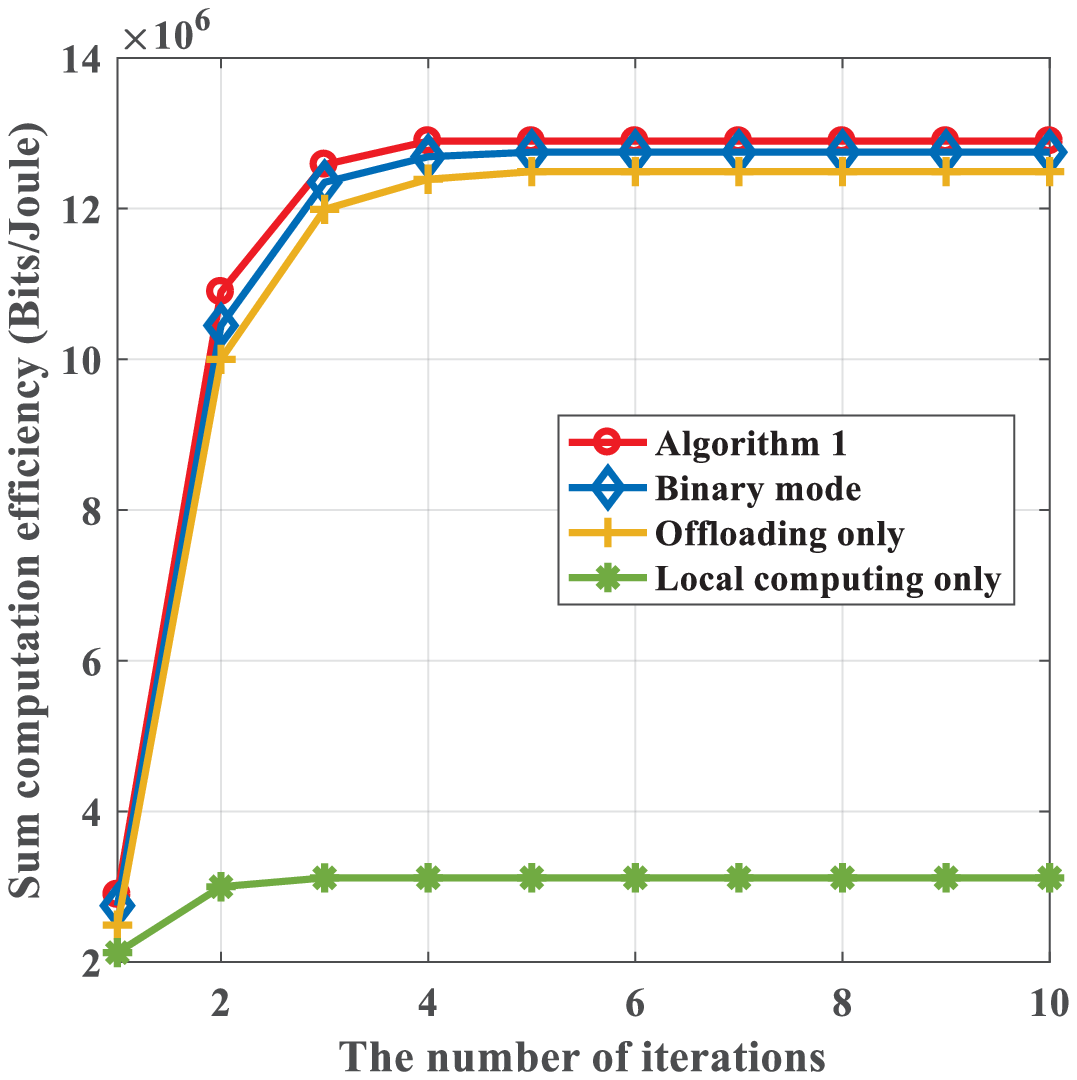}
\end{minipage}
}
\subfigure[]{
\begin{minipage}{2.25 in}
\centering
\includegraphics[width=2.3 in]{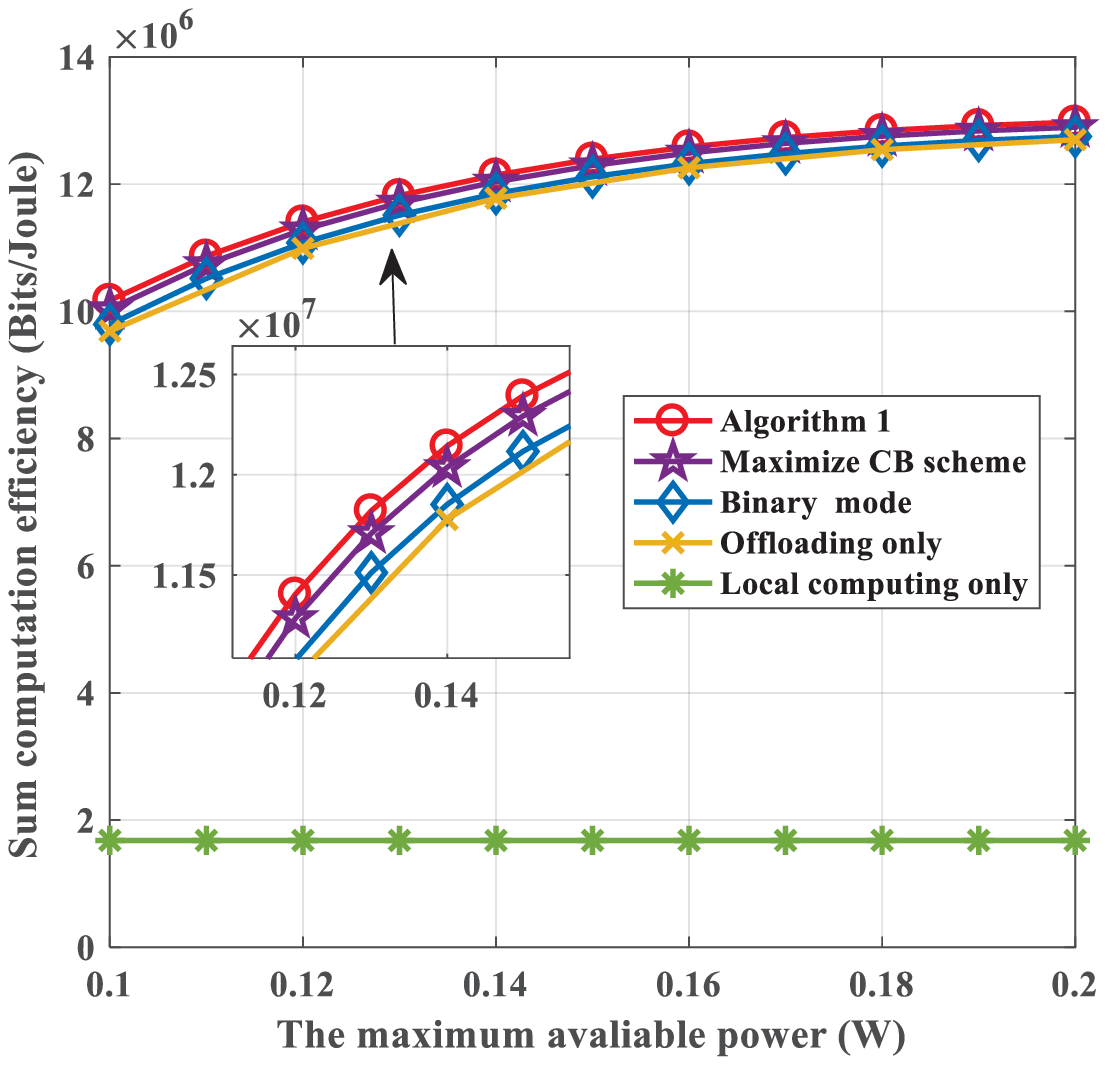}
\end{minipage}
}
\subfigure[]{
\begin{minipage}{2.25 in}
\centering
\includegraphics[width=2.3 in]{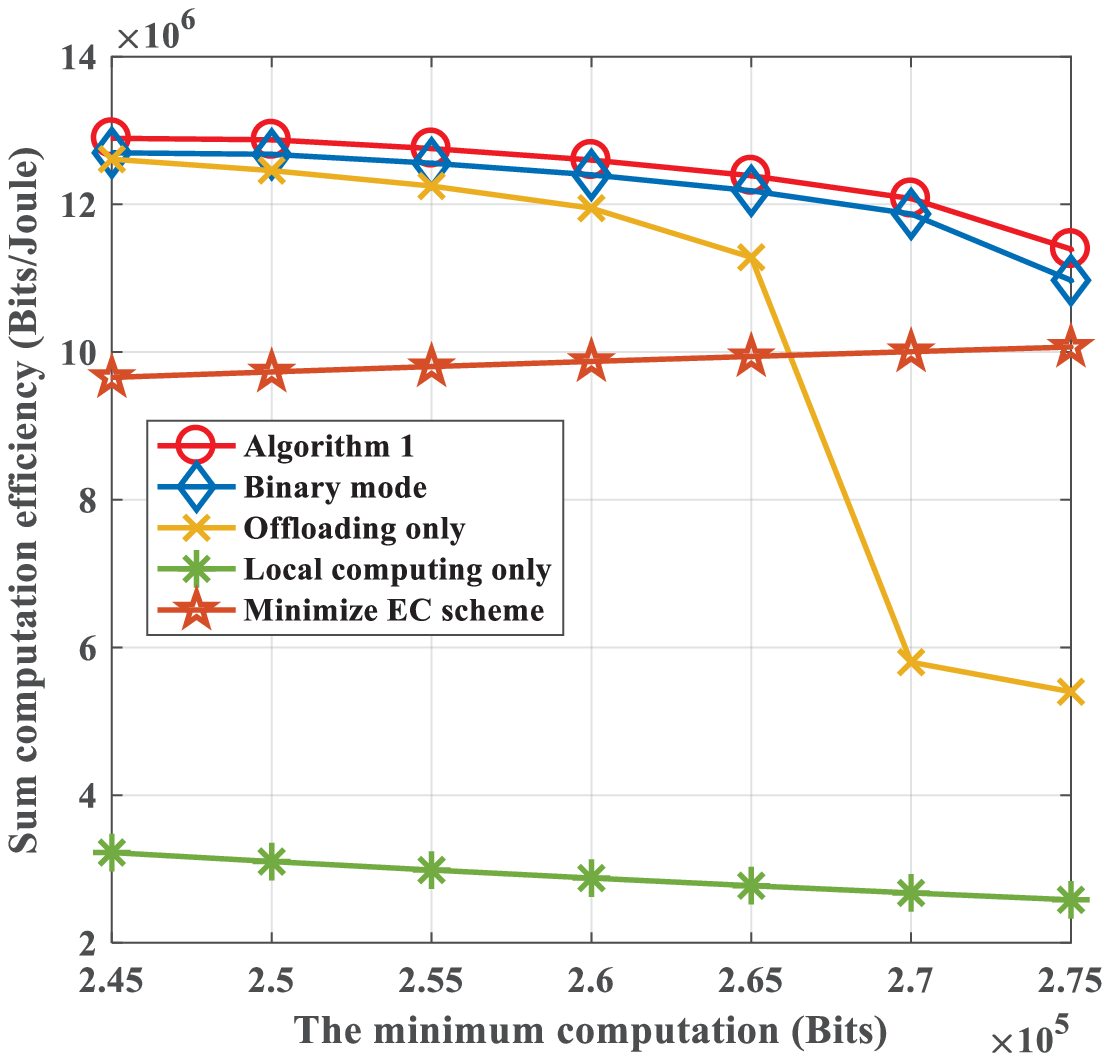}
\end{minipage}
}
\caption{(a)~Convergence of the proposed algorithm for different schemes.; (b)~The sum CE versus the maximum available energy; (c)~The sum CE versus the minimum data requirement.}
\end{figure*}

Fig. 3 illustrates the weighted sum CE achieved with those four schemes versus the minimum computed bits requirement. As shown in Fig. 3, the CE do not increase with the minimum computation bits. It clearly indicates that the tradeoff between the computed bits and the CE exists. It is also shown that the performance of the proposed schemes is superior to that of three benchmark schemes with respect to the CE. In particular, the CE achieved under the EC scheme is smaller than that obtained under the proposed CE maximization scheme. The reason is that the energy consumption minimization scheme cannot consume more energy to increase the achievable computation bits as long as the minimum computation bits requirement is satisfied. Therefore, the maximum CE cannot be achieved under the energy consumption minimization scheme. This also indicates that our framework can achieve a good tradeoff between the consumed energy and  the computation bits.
\section{Conclusions}
The weighted sum CE maximization problems were studied in an OFDMA-based MEC system. The subchannel, tranmit power and local CPU frequency were jointly optimized to maximized the weighted sum CE. By exploiting the sum-of-ratio structure of the objective function, we proposed an efficiently iterative algorithm to solve the formulated problems. Simulation results demonstrated the efficiency of our proposed algorithm. It was also shown that our proposed resource allocation strategy has a larger CE compared to the benchmark schemes. Moreover, it shows that the CE obtained by the partial offloading mode is better than that achieved in the binary offloading mode. Furthermore, the tradeoffs between the computation efficiency and the number of computation bits optimization framework are revealed.

\end{document}